\documentclass[osajnl,preprint,showpacs,superscriptaddress,12pt]{revtex4-1} 

\usepackage{amsmath,amssymb,graphicx}

\begin{document}

\title{Phase state and related nonlinear coherent states}

\author{F. Soto-Eguibar}

\author{B. M. Rodr\'{\i}guez-Lara}
\email{Corresponding author: bmlara@inaoep.mx}

\author{H. M. Moya-Cessa}
\affiliation{Instituto Nacional de Astrof\'{i}sica, \'{O}ptica y Electr\'{o}nica \\ Calle Luis Enrique Erro No. 1, Sta. Ma. Tonantzintla, Pue. CP 72840, M\'{e}xico}

%%%%%%%%%%%%%%%%%%%%%%%%%%%%%%%%%%%%%%%%%%%%%%%%%%%%%%%%%%%%%%%%%%%%%%%%%%%%%%%%%%%%%%%%
\begin{abstract}
We cast the phase state as a $SU(1,1)$ nonlinear coherent state to support the idea that the $SU(1,1)$ representation of the electromagnetic field may be helpful in some instances and to bring forward that it may relate to the phase state problem. 
We also construct nonlinear coherent states related to the exponential phase operator and provide their corresponding nonlinear annihilation operators.
Finally, we discuss the propagation of classical fields through arrays of coupled waveguides that are solved through the use of nonlinear coherent states of $SU(1,1)$ or the exponential phase operator.
\end{abstract}
%%%%%%%%%%%%%%%%%%%%%%%%%%%%%%%%%%%%%%%%%%%%%%%%%%%%%%%%%%%%%%%%%%%%%%%%%%%%%%%%%%%%%%%%

%\pacs{42.50.Ct, 42.50.-p, 42.50.Pq, 42.50.Dv}
%\ocis{ (270.0270) Quantum optics;  (270.5580) Quantum electrodynamics; (230.7370) Waveguides;}

\maketitle

%%%%%%%%%%%%%%%%%%%%%%%%%%%%%%%%%%%%%%%%%%%%%%%%%%%%%%%%%%%%%%%%%%%%%%%%%%%%%%%%%%%%%%%%
\section{Introduction} \label{sec:S1}
%%%%%%%%%%%%%%%%%%%%%%%%%%%%%%%%%%%%%%%%%%%%%%%%%%%%%%%%%%%%%%%%%%%%%%%%%%%%%%%%%%%%%%%%

The idea of a quantum phase operator is an old problem in quantum mechanics \cite{Carruthers1968p411}. 
In 1926, London found out that there is no quantum mechanics Hermitian operator that corresponds to the phase variable in classical mechanics \cite{London1926p915} and in its place proposed an exponential phase operator, $\widehat{e^{i \phi}}$, with problems of its own \cite{London1927p193}.
The idea of a phase operator came to the field of quantum optics when Dirac approached the quantization of the electromagnetic field via angle variables \cite{Dirac1927p243} despite the problems pointed out by London and, after a while,  Susskind and Glogower rediscovered the exponential phase operator \cite{Susskind1964p49}.

Here our interest is twofold.
First, there is an effort in theoretical physics to bring forward bosons as multimode coherent states of the universal covering group of $SU(1,1)$ \cite{Celeghini1995p239,Celeghini1996p1625,Celeghini1998p3424,Celeghini1999p2909}.
Such an approach may not simplify the problems found in quantum optics, which are well developed through the Heisenberg-Weyl group provided by the number, creation and annihilation operators, but we will show in the following that it is possible to cast the phase state as a generalized $SU(1,1)$ coherent state based on the Lie algebraic representation of quantum phase and number operators \cite{Kastrup2003p975,Rasetti2004p479,Kastrup2006p052104}.
We do not pretend to touch upon the phase problem, but our approach may provide further support to the $SU(1,1)$ formalism and may open a new avenue to approach some quantum optics problems.
Second, coherent states have proved useful in describing the quantum electromagnetic field since their introduction to quantum optics by Sudarshan \cite{Sudarshan1963p277} and Glauber \cite{Glauber1963p2766}; we owe their inception as minimum uncertainty product states in quantum mechanics to Schr\"odinger \cite{Schrodinger1926p664}.
Some sets of nonlinear coherent states of the field have been brought forward in quantum optics recently \cite{Sivakumar2002p6755,Recamier2008p673,delosSantosSanchez2012p015502,SantosSanchez2013p375303,RomanAcheyta2014p38} and, here, we want to provide a couple of nonlinear coherent states related to the exponential phase operator \emph{a l\`{a}} Perelomov \cite{Perelomov1972p222} and relate them to operators that are diagonal in those nonlinear coherent bases  \emph{a l\`{a}} Barut and Girardello \cite{Barut1971p41}.
Finally, we bring forward the propagation of classical light in arrays of coupled waveguides as an example of how these nonlinear coherent states provide a simple solution to their classical optics analogues.

%%%%%%%%%%%%%%%%%%%%%%%%%%%%%%%%%%%%%%%%%%%%%%%%%%%%%%%%%%%%%%%%%%%%%%%%%%%%%%%%%%%%%%%%
\section{Phase state as a $SU(1,1)$ coherent state} \label{sec:S2}
%%%%%%%%%%%%%%%%%%%%%%%%%%%%%%%%%%%%%%%%%%%%%%%%%%%%%%%%%%%%%%%%%%%%%%%%%%%%%%%%%%%%%%%%

We first want to show that the phase state in terms of Fock states \cite{Loudon1973,Schleich2001}, 
\begin{eqnarray}
\vert \phi \rangle \equiv \frac{1}{\sqrt{2 \pi}} \sum_{j=0}^{\infty} e^{i \phi \left(j + \frac{1}{2} \right)} \vert j \rangle,
\end{eqnarray} 
can be cast in the form of a generalized $SU(1,1)$ coherent state.
First, let us define the $SU(1,1)$ group elements, $\hat{K}_{0} = \hat{n} + k$,  $\hat{K}_{+} = \hat{a}^{\dagger} \sqrt{\hat{n} + 1}$ and  $\hat{K}_{-} = \sqrt{\hat{n} + 1} ~\hat{a}$ with Bargmann parameter $k=1/2$, in terms of the creation (annihilation) operators, $\hat{a}^{\dagger}$ ($\hat{a}$),  such that $\left[ \hat{K}_{0} , \hat{K}_{\pm} \right] = \pm \hat{K}_{\pm}$ and $\left[ \hat{K}_{+} , \hat{K}_{-} \right] = -2 \hat{K}_{0}$ \cite{Buck1981p132}.
Realizing that $\hat{K}_{0} \vert k, n \rangle = (n + k) \vert k, n \rangle$,  $\hat{K}_{+}^{j} \vert k, 0 \rangle = j! \vert k, j \rangle$ and obviating the Bargmann parameter, $\vert n \rangle \equiv \vert k, n \rangle$, leads us to write:
\begin{eqnarray}
\vert \phi \rangle = \frac{1}{\sqrt{2 \pi}}  e^{i \phi \hat{K}_{0}} e^{\hat{K}_{+}} \vert 0 \rangle.
\end{eqnarray} 
Now, we can use the normal to antinormal ordering expressions (2.16) - (2.20) in \cite{Ban1993p1348},
\begin{eqnarray}
e^{A_{+} \hat{K}_{+}} e^{\ln A_{0}~ \hat{K}_{0}} e^{A_{-} \hat{K}_{-}} = e^{B_{-} \hat{K}_{-}} e^{\ln B_{0} ~\hat{K}_{0}} e^{B_{+} \hat{K}_{+}}
\end{eqnarray}
with 
\begin{eqnarray}
A_{\pm} &=& \frac{ B_{\pm} B_{0} }{ 1 - B_{+} B_{0} B_{-} }, \\
A_{0} &=& \frac{  B_{0} }{ \left( 1 - B_{+} B_{0} B_{-} \right)^{2} }, \\
B_{\pm} &=& \frac{ A_{\pm}  }{ 1 - A_{+} A_{-} }, \\
B_{0} &=& \frac{ \left( A_{0} - A_{+} A_{-} \right)^{2}   }{ A_{0} }.
\end{eqnarray}
In our case the parameters are $B_{-}=0$, $B_{0}= e^{i \phi}$, $B_{+}=1$ and we have used the fact that $e^{\xi \hat{K}_{-}} \vert 0 \rangle = \vert 0 \rangle$ to reach our goal and write a phase state as a generalized $SU(1,1)$ coherent state in a Perelomov-like form,
\begin{eqnarray}
\vert \phi \rangle = \frac{1}{\sqrt{2 \pi}}   e^{e^{i \phi}\hat{K}_{+}} e^{i \phi \hat{K}_{0}} e^{-e^{-i \phi}\hat{K}_{-}} \vert 0 \rangle.
\end{eqnarray} 
Furthermore, we already know that the phase state is an eigenstate of the exponential phase operator.
Casting it in terms of $SU(1,1)$ elements allows us to see it as a nonlinearly deformed annihilation operator whose eigenstate is the phase state,
\begin{eqnarray}
\widehat{e^{i \phi}} \vert \phi \rangle &=& \frac{1}{\sqrt{\hat{n} + 1}} \hat{a} \vert \phi \rangle, \\
&=& \frac{1}{\hat{K}_{0} + \frac{1}{2}} \hat{K}_{-} \vert \phi \rangle, \\
&=& e^{i \phi} \vert \phi \rangle,
\end{eqnarray}
which makes the phase state a Barut-Girardello nonlinear coherent state of the exponential phase operator; i.e., it can be seen as a deformed annihilation operator $ \hat{C}=f(\hat{A}_{0}) \hat{A}_{-}$ with $f(\hat{A}_{0}) = f(\hat{n})=\left( \hat{n} + 1\right)^{-1/2}$ and $\hat{A}_{-} = \hat{a}$ for the Heisenberg-Weyl group  or $f(\hat{A}_{0})= f(\hat{K}_{0})= \left(\hat{K}_{0} + 1/2 \right)^{-1}$ and $\hat{A}_{-}= \hat{K}_{-}$ for $SU(1,1)$.

%%%%%%%%%%%%%%%%%%%%%%%%%%%%%%%%%%%%%%%%%%%%%%%%%%%%%%%%%%%%%%%%%%%%%%%%%%%%%%%%%%%%%%%%
\section{London nonlinear coherent states}
%%%%%%%%%%%%%%%%%%%%%%%%%%%%%%%%%%%%%%%%%%%%%%%%%%%%%%%%%%%%%%%%%%%%%%%%%%%%%%%%%%%%%%%%
Now we want to construct nonlinear coherent states related to the exponential phase operator proposed originally by London; thus, we have christened them London nonlinear coherent states for short.
Let us start with the Barut-Girardello coherent state for the non-compact operator $\hat{K}_{-}$ of $SU(1,1)$ with Bargmann parameter $k=1/2$ \cite{Puri2001,Chakrabarti2003p287}, 
\begin{eqnarray}
\vert \alpha_{BG} \rangle = \frac{1}{\sqrt{I_{0}(2 \vert \alpha \vert)}} \sum_{j=0}^{\infty} \frac{\alpha^{j}}{j!} \vert j \rangle,
\end{eqnarray}
such that $\hat{K}_{-} \vert \alpha_{BG} \rangle = \alpha \vert \alpha_{BG} \rangle$, where we are keeping all the definitions from the previous section and the function $I_{n}(x)$ is the $n$th modified Bessel function of the first kind.
If we recover the action of the exponential phase operators, $\hat{V} \equiv \widehat{e^{i \phi}} = \left( \hat{n} + 1 \right)^{-1/2} \hat{a}$ and $\hat{V}^{\dagger} \equiv \widehat{e^{-i \phi}} =  \hat{a}^{\dagger} \left( \hat{n} + 1 \right)^{-1/2}$, over Fock states, $\hat{V} \vert n \rangle = \vert n-1 \rangle$ and $\hat{V}^{\dagger} \vert n \rangle = \vert n +  1 \rangle$, it is straightforward to write this Barut-Girardello coherent states as Perelomov-like nonlinear coherent states, 
\begin{eqnarray}
\vert \alpha_{BG} \rangle =  \frac{1}{\sqrt{I_{0}(2 \vert \alpha \vert)}} e^{\alpha \hat{V}^{\dagger}} e^{-\alpha^{\ast} \hat{V}} \vert 0 \rangle. \label{eq:Barut}
\end{eqnarray}

Now, due to the problems arising from the right-unitarity of the exponential phase operators, $\hat{V} \hat{V}^{\dagger} = \hat{1}$ but $\hat{V}^{\dagger} \hat{V} = \hat{1} - \vert 0 \rangle \langle 0 \vert$, trying to write Eq. (\ref{eq:Barut}) as a nonlinear displacement operator acting on the vacuum is beyond our current scope but we can bring forward another nonlinear coherent state \emph{a l\`a} Perelomov related to the exponential phase operator \cite{LeonMontiel2011p133},
\begin{eqnarray}
\vert \alpha \rangle &=& e^{\alpha \left( \hat{V}^{\dagger} - \hat{V} \right)} \vert 0 \rangle, \quad \alpha \in \mathbb{R}, \label{eq:NLLSD} \\
&=& \frac{1}{\alpha} \sum_{j=0}^{\infty} \left( j + 1 \right) J_{j+1}(2 \alpha) \vert j \rangle, \label{eq:ExpVdmV}
\end{eqnarray}
which is a Barut-Girardello coherent state, $\hat{C}_{\alpha} \vert \alpha \rangle = \alpha \vert \alpha \rangle$, of the $\alpha$-deformed annihilation operator,
\begin{eqnarray}
\hat{C}_{\alpha} &=& \frac{\alpha J_{\hat{n}+1}(2 \alpha)}{\left( \hat{n} + 2 \right) J_{\hat{n}+2}(2 \alpha)} \sqrt{\hat{n}+1} ~\hat{a} , \\
&=& \frac{\alpha J_{\hat{K}_{0}+\frac{1}{2}}(2 \alpha)}{\left( \hat{K}_{0} + \frac{3}{2} \right) J_{\hat{K}_{0}+\frac{3}{2}}(2 \alpha)} \hat{K}_{-} , 
\end{eqnarray} 
where the function $J_{m}(x)$ is the $m$th Bessel function of the first kind.
Note that while Eq. (\ref{eq:NLLSD}) allows for any real value of parameter $\alpha$, the definition of the related operator seems to point that the coherent parameter $\alpha$ should never be half a root of a Bessel function,  $J_{n+2}(2 \alpha) \ne 0$ for all $n \ge 0$; this occurrence may be similar to Eq. (\ref{eq:ExpVdmV}), where it may appear that the coherent parameter must not be zero but such a value is allowed by Eq. (\ref{eq:NLLSD}) or be a topological issue related to the definition of generalized coherent states \cite{rasetti75, rasetti85, rasetti89}.

%%%%%%%%%%%%%%%%%%%%%%%%%%%%%%%%%%%%%%%%%%%%%%%%%%%%%%%%%%%%%%%%%%%%%%%%%%%%%%%%%%%%%%%%
\section{Classical optics examples}
%%%%%%%%%%%%%%%%%%%%%%%%%%%%%%%%%%%%%%%%%%%%%%%%%%%%%%%%%%%%%%%%%%%%%%%%%%%%%%%%%%%%%%%%

The quintessential examples involving the $SU(1,1)$ group belong to quantum optics.
It is well known that the degenerate parametric oscillator preserves \cite{Gerry1985p2721} and generates \cite{Wodkiewicz1985p458} $SU(1,1)$ coherent and squeezed states, in that order.
It has also been theoretically proposed \cite{Schleich1989p7405} and experimentally shown \cite{Breitenbach1997p2207} that the phase state probability of a highly squeezed state shows a bifurcation as a function of the squeezing parameter.
A theoretical program emerged to approach linear dissipative processes in quantum optical systems related to phase modulation and photon echo \cite{Ban1992p3213}.
Even purely theoretical models such as the Buck-Sukumar model \cite{Buck1981p132,Buzek1989p1151} and the anharmonic oscillator \cite{Gerry1987p2146,Buzek1990p393} have shown the benefits of using the $SU(1,1)$ formalism in quantum optics although care must be exerted depending on the particular circumstances \cite{Lo1999p557,Ng2000p463,RodriguezLara2014}.

Here we are interested in providing classical optics examples in which the use of the $SU(1,1)$ or London nonlinear coherent states simplifies the problem of describing the propagation of a classical field through a photonic lattice.
Arrays of waveguides have provided a classical simulator of quantum and relativistic physics \cite{Longhi2009p243,Longhi2011p453}. 
In particular, some of us have shown classical analogues to quantum coherent states in one-dimensional photonic lattices \cite{Keil2011p103601,RodriguezLara2011p053845,PerezLeija2012p013848} and have used the $SU(1,1)$ group to propose isospectral arrays of coupled waveguides \cite{ZunigaSegundo2014p987}.
First, allow us to consider a semi-infinite lattice composed of identical waveguides and described by the effective differential equation set,
\begin{equation}
-i \frac{d}{dz}\mathcal{E}_{j}(z) = \left[ \sqrt{(j+1)(j+2)} \mathcal{E}_{j+1} + \sqrt{j(j+1)} \mathcal{E}_{j-1} \right],
\end{equation}
where the field amplitude at the $j$th waveguide is given by $\mathcal{E}_{j}$ and  $\mathcal{E}_{- \vert j \vert} =0$. 
This system is equivalent to that studied in \cite{PerezLeija2011p1833} with parameter $\chi=1$ and it is straightforward to show that this is equivalent to the Schr\"odinger-like equation with $\vert E \rangle = \sum_{j} \mathcal{E}_{j} \vert k,j\rangle$ and Bargmann parameter $k=1/2$,
\begin{eqnarray}
i \frac{d}{dz} \vert E \rangle =  \left( \hat{K}_{+} + \hat{K}_{-} \right) \vert E \rangle.
\end{eqnarray}
We can calculate the impulse function, $\mathcal{I}_{m,n}$ which is the field at the $m$th waveguide given that the initial field impinged just the $n$th waveguide, as a projected $SU(1,1)$ coherent state by use of well-known formulas. 
In this case we can use Eq. (3.1) in \cite{Ban1993p1348},
\begin{eqnarray}
\vert k, \alpha \rangle &=&e^{ \alpha \hat{K}_{+} - \alpha^{\ast} \hat{K}_{-}  } \vert k,0 \rangle ,\\
&=& \left( 1 - \vert \mu \vert^{2} \right)^{k} \sum_{m=0}^{\infty} \sqrt{\frac{\Gamma(2k + m)}{m! \Gamma(2k)}} ~\mu^{m} \vert k,m \rangle,
\end{eqnarray}
where $\mu = \left(\alpha / \vert \alpha \vert \right) \tanh \vert \alpha \vert$, with $z= i \alpha$ and $k=1/2$ to obtain
\begin{eqnarray}
\mathcal{I}_{m,0}(z) &=& \langle m \vert e^{i  z \left( \hat{K}_{+} + \hat{K}_{-} \right)  } \vert 0 \rangle, \\
&=& \mathrm{sech}z \left(i ~\mathrm{tanh}~z \right)^{m},
\end{eqnarray}
where we have obviated, again, the Bargmann parameter in the notation for the basis.
Thus, the classical field distribution for a starting field impinging the zeroth waveguide is identical to the distribution of a $SU(1,1)$ coherent state $\vert k, i z \rangle =  e^{i  z \left( \hat{K}_{+} + \hat{K}_{-} \right)  } \vert j, 0 \rangle$.
Note that the increasing coupling strength between waveguides will mean that at some point second-, third- and higher-order neighbor couplings should be taken into account and, thus, the experimental realization of this example is not trouble-free.

An example involving London coherent operators is produced by a semi-infinite array of identical waveguides that are homogeneously coupled \cite{Makris2006p036616,LeonMontiel2011p349}, 
\begin{eqnarray}
-i \frac{d}{dz}\mathcal{E}_{j}(z) =  \mathcal{E}_{j+1} +  \mathcal{E}_{j-1} ,
\end{eqnarray}
leading to the Schr\"odinger-like equation
\begin{eqnarray}
- i \frac{d}{dz} \vert E \rangle =  \left( \hat{V}^{\dagger} + \hat{V} \right) \vert E \rangle.
\end{eqnarray}
Again, it is straightforward to write
\begin{eqnarray}
\mathcal{I}_{m,0}(z) &=& \langle m \vert e^{i  z \left( \hat{V}^{\dagger} + \hat{V} \right)  } \vert 0 \rangle, \\
&=& \frac{1}{z} i^{m} \left( m +1 \right) J_{m+1}(2z),
\end{eqnarray}
where we have used Eq. (\ref{eq:ExpVdmV}) and the fact that $e^{- i \pi \hat{n} /2 } e^{i \alpha \left( \hat{V}^{\dagger} + \hat{V} \right)} e^{i \pi \hat{n} /2 } = e^{ \alpha \left( \hat{V}^{\dagger} - \hat{V} \right)}$ with $\alpha \in \mathbb{R}$.
Here the field amplitude distribution for an initial field impinging just the zeroth waveguide corresponds to the distribution of a London nonlinear coherent state defined by $\vert iz \rangle = e^{i  z \left( \hat{V}^{\dagger} + \hat{V} \right)  } \vert 0 \rangle$.
In this case the experimental realization only has to take into account that the length and size of the photonic lattice should keep light far away from the last waveguide. 
In both cases, the distribution is that of a $SU(1,1)$ or London displaced number state when the field impinges the $n$th waveguide with $n \ne 0$.

%%%%%%%%%%%%%%%%%%%%%%%%%%%%%%%%%%%%%%%%%%%%%%%%%%%%%%%%%%%%%%%%%%%%%%%%%%%%%%%%%%%%%%%%
\section{Conclusions} \label{sec:S6}
%%%%%%%%%%%%%%%%%%%%%%%%%%%%%%%%%%%%%%%%%%%%%%%%%%%%%%%%%%%%%%%%%%%%%%%%%%%%%%%%%%%%%%%%
We have shown that describing the quantum electromagnetic field via the $SU(1,1)$ algebra leads to a representation of the phase state as a generalized coherent state. 
The use of this formalism may simplify the work needed to study quantum optical systems such as anharmonic oscillators, degenerate parametric oscillators or two-mode couplers in phase representation.
We took the opportunity, arisen from describing the phase state as a nonlinear coherent state, to introduce other nonlinear coherent states related to the exponential phase operator.
We showed that the Barut-Girardello coherent state for $\hat{K}_{-}$ can be seen as a Perelomov-like coherent state related to the exponential phase operators.
Also, we brought forward the operators that have as proper states the vacuum displaced  via exponential phase operators.
All the nonlinear coherent states described here were cast in both a Barut-Girardello eigenvalue relation and Perelomov-like form.
Finally, we discussed the propagation of classical light through two arrays of coupled waveguides where the impulse function can be given in closed form via the phase formalism.

%%%%%%%%%%%%%%%%%%%%%%%%%%%%%%%%%%%%%%%%%%%%%%%%%%%%%%%%%%%%%%%%%%%%%%%%%%%%%%%%%%%%%%%%
%\section*{Acknowledgments}
%%%%%%%%%%%%%%%%%%%%%%%%%%%%%%%%%%%%%%%%%%%%%%%%%%%%%%%%%%%%%%%%%%%%%%%%%%%%%%%%%%%%%%%%

%%%%%%%%%%%%%%%%%%%%%%%%%%%%%%%%%%%%%%%%%%%%%%%%%%%%%%%%%%%%%%%%%%%%%%%%%%%%%%%%%%%%%%%%
%\bibliographystyle{osajnl}
%\bibliography{D:/ExternalHD/Bibliography/references}

\end{document}